\documentclass[english,prl,twocolumn,showpacs,amsmath,amssymb,superscriptaddress,floatfix]{revtex4}
\usepackage[T1]{fontenc}
\usepackage[latin1]{inputenc}
\usepackage{graphicx}

\makeatletter

\usepackage{bm}
\usepackage{pifont}
\usepackage{babel}
\makeatother

\begin{document}

%\preprint{}

\title{Measuring the complex admittance of a carbon nanotube double quantum dot}

\author{S.J. Chorley}
\author{J. Wabnig}
\author{Z.V. Penfold-Fitch}
\author{K.D. Petersson}
\altaffiliation[Current address: ]{Department of Physics, Princeton University, Princeton, NJ 08544, USA}
\author{J. Frake}
\author{C.G. Smith}
\author{M.R. Buitelaar}\email{mrb51@cam.ac.uk}
\affiliation{Cavendish Laboratory, University of Cambridge, Cambridge CB3 0HE, United Kingdom}%

\begin{abstract}
We investigate radio-frequency (rf) reflectometry in a tunable
carbon nanotube double quantum dot coupled to a resonant circuit. By
measuring the in-phase and quadrature components of the reflected rf
signal, we are able to determine the complex admittance of the
double quantum dot as a function of the energies of the
single-electron states. The measurements are found to be in good
agreement with a theoretical model of the device in the incoherent
limit. Besides being of fundamental interest, our results present an
important step forward towards non-invasive charge and spin state
readout in carbon nanotube quantum dots.
\end{abstract}


% 73.63.Fg Nanotubes
% 73.63.Kv Quantum dots
% 73.23.Hk Coulomb blockade; single-electron tunneling
% 03.67.Lx Quantum computation architectures and implementations

% 71.70.Ej Spin-orbit coupling, Zeeman and Stark splitting, Jahn-Teller effect
% 73.21.La Quantum dots

%Use showkeys class option if keyword display desired

%\keywords{carbon nanotube, quantum dot, rf-reflectometry, capacitance}

\maketitle

An important requirement in any quantum information processing
scheme is fast manipulation and readout of the quantum system in
which the quantum information is encoded. This requires an
understanding of the response of the quantum system at finite
frequencies which, in the case of an electronic device, involves an
understanding of its complex admittance \cite{Buttiker1,Buttiker2}.
Of particular interest in the context of quantum information
processing are double quantum dots which are widely used to define
charge and spin qubits \cite{Loss}. However, while double quantum
dots have been investigated in detail over the last decade,
experiments to measure and analyze their complex admittance have not
yet been performed and this topic has only recently been addressed
theoretically \cite{Cottet}. The admittance of quantum dots at
finite frequencies is non-trivial as exemplified by recent
experiments on single quantum dots \cite{Gabelli, Delbecq}. The
physics is even richer for double quantum dots as internal charge
dynamics, i.e. charge transfer between the quantum dots, has to be
taken into account. However, the dependence of the admittance on the
internal charge dynamics also provides a route towards charge and
spin state readout \cite{Petersson}.

In this work we present a detailed experimental study of the complex
admittance of a carbon nanotube double quantum dot which is measured
using rf reflectometry techniques. The measurements are compared
with a theoretical model of the device where we use a density matrix
approach to calculate the double quantum dot admittance. The good
quantitative agreement between the experimental and theoretical
results allows us to determine the effective conductance and
susceptance of the double dot as a function of the energies of the
single-electron states. Our measurements thus present a first
quantitative analysis of the complex admittance of a double quantum
dot. The demonstrated technique also provides the basis for a simple
and fast detection scheme for charge and spin state readout in
carbon nanotubes - a material with considerable potential for
spin-based quantum information processing \cite{Kuemmeth,
Galland,Churchill1, Churchill2,Jespersen,Chorley} - without the need
for a separate charge detector \cite{Biercuk}.

\begin{figure}
\includegraphics[width=85mm]{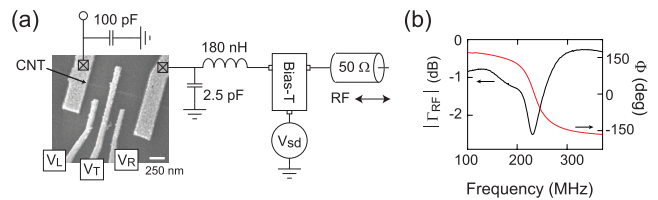}
\caption{\label{Fig1}\textbf{(a)} (color online) Schematic of the
carbon nanotube double quantum dot and resonant circuit. The device is connected to a 50 $\Omega$ transmission line
for the rf reflectometry measurements. A dc signal is applied via a bias-tee. \textbf{(b)} Measured amplitude and phase response for the resonant circuit
relative to the transmission line background.}
\end{figure}

The device we consider is a carbon nanotube grown by chemical vapour
deposition on degenerately doped Si terminated by 300 nm SiO$_2$,
see Fig.~1(a). The nanotube is contacted by Au source and drain
electrodes which form the outer tunnel barriers of the quantum dots.
A capacitively coupled top gate, separated from the nanotube by
$\sim$ 3 nm AlO$_x$, is used to define a tunable coupling between
the dots while two plunger gates vary the energies of the dots. The
nanotube device is embedded in a resonant circuit consisting of a
parasitic capacitance $C=2.5$ pF and on-chip inductor $L=180$ nH
\cite{capacitance}. The circuit has a resonance frequency $f_0 \sim
236$ MHz and loaded quality factor $Q \sim 1/Z_0 \sqrt{L/C} \sim
5.4$, where $Z_0 = 50$ $\Omega$ is the characteristic impedance of
the transmission lines, see Fig.~1(b). We note that higher quality
factors ($Q \sim 30$) were readily obtained on nanotube devices
grown on undoped Si and quartz substrates. A highly doped Si
substrate is used here because of its convenience as a back gate.

\begin{figure}
\includegraphics[width=85mm]{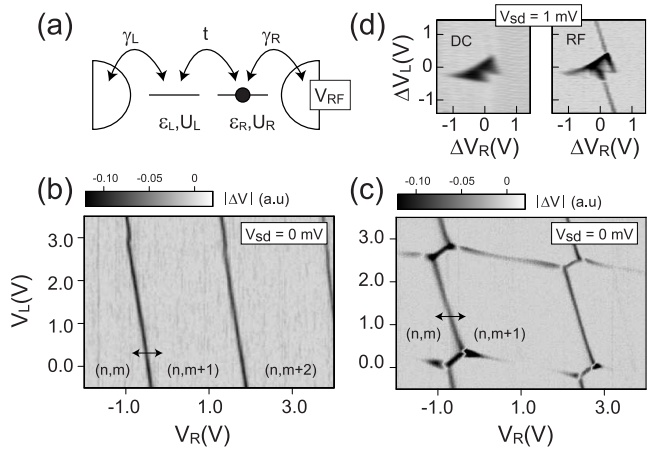}
\caption{\label{Fig2}\textbf{(a)} Schematic of the
carbon nanotube double quantum dot device. The rf signal is applied to the right electrode. \textbf{(b,c)} Demodulated response of the
resonant circuit as a function of $V_L$ and $V_R$. In each plot the
background signal measured inside a stable charge region is subtracted. The top gate voltage
is set to $V_t = 0$ V. The back gate voltages are $V_{bg} = 5$ V and $V_{bg} = -0.18$ V in (b) and (c), respectively. \textbf{(d)} Measured rf signal (right) and
dc current (left) for a triple point pair in the presence of an applied bias
$V_{sd} = 1$ mV. The full scale of the dc current represents 150 pA.}
\end{figure}

The nanotube device is characterized by dc transport measurements and rf reflectometry in a dilution refrigerator with
a base temperature $T \sim 40$ mK. The dc signal is applied via a bias-tee while the rf excitation is
directed to the source electrode of the double quantum dot through the coupled port of a directional
coupler, connected to the sample holder via a stainless-steel
semi-rigid coaxial line, see Fig.~1(a). The reflected signal is sent back via a
cryogenic preamplifier which is thermally anchored at 4 K, followed
by room temperature amplification and demodulation by mixing with the reference signal. We drive the resonator at its resonant frequency with an amplitude at the double dot $V_{RF} \sim 10$ $\mu$eV. Depending on the energies of the quantum dots, the oscillating
potential on the source electrode may induce charge transfer (tunneling) between the electrodes and the quantum dots and/or redistribution of charge between the two dots. The resulting oscillating current will generally
have both out-of-phase and in-phase components with respect to the driving voltage which give rise to a resonance frequency shift and damping of the resonator. The response of the double quantum dot is thus characterized by a complex admittance which can be deduced by measuring the phase and amplitude of the reflected rf signal.

We first determine the various energy scales of the double quantum dot by measuring the dc and rf response as a function of the side-gate voltages
$V_L$ and $V_R$ which modulate the left and right quantum dot energies, respectively, see Fig.~2. For simplicity in-phase and out-of-phase components of the rf response are shown together here (i.e. the demodulated signal is sensitive to both amplitude and phase). The nanotube device is fully tunable by the top $V_t$ and back gate $V_{bg}$ voltages. The stability diagram of Fig.~2(b), for example, illustrates the situation where the two quantum dots are fully decoupled: no dc current could be detected but a strong rf response is observed at charge transitions of the right quantum dot. For more positive back gate voltages, the stability diagram evolves into a honeycomb pattern that is characteristic of the double quantum dot investigated here. At finite bias $V_{sd} = 1$ mV, bias triangles are observed at the triple points in both the dc and rf response, see Fig.~2(d), which allows us to extract the characteristic energy scales of the double quantum dot. We obtain charging energies $U_L \sim 6.5$ meV and $U_R \sim 5$ meV for the left and
right quantum dots, respectively, and an inter-dot electrostatic coupling energy $U'=0.6$ meV. The analysis also allows us to deduce
the various geometric capacitances of the device \cite{Wiel}. We did not observe any obvious four-fold periodicity
in the stability diagrams, an indication that the orbital degeneracy of the nanotubes is broken \cite{Liang, Buitelaar}.

To determine the complex admittance of the nanotube device we
measured the in-phase and quadrature components of the reflected
signal for the double quantum dot which allows us to extract the
amplitude and phase information, see Fig.~3(a) and (b). The observed
phase shifts $\Delta \Phi$, relative to the phase measured inside a
stable charge region, are strongest at the $(n,m)-(n,m+1)$ charge
transition, with a signal of about half the strength observed at the
internal $(n+1,m)-(n,m+1)$ charge transition. The amplitude shifts
on the other hand are concentrated around the triple points. Using
standard circuit analysis \cite{Pozar}, the measured phase signal at
resonance can be related to a change in capacitance as $\Delta
\Phi/\Delta C \sim 2Q/C$, where $C = 2.5$ pF for the circuit
considered here. At the $(n,m)-(n,m+1)$ charge transition, for
example, the measured phase shift of $\sim 0.18$ degrees implies
$\Delta C = 0.74$ fF. Note that this is several orders of magnitude
larger than the geometric capacitances of carbon nanotube quantum
dots which are typically in the aF range. The amplitude modulation
is related to the double dot resistance $R$ via $\Delta |\Gamma|
/|\Gamma| = 2Q^2Z_0/R$. The measured damping $\Delta |\Gamma|
/|\Gamma| \sim 0.2 \%$ at the triple points therefore implies $R
\sim 1.5$ M$\Omega$ which is in agreement with dc conductance
measurements.

In the following we compare the experimental results with a theoretical model of the device, the full results of which
are shown in Fig.~3(c) and (d) which show the real ($R^{-1}_{eff}$) and imaginary ($C_{eff}$) part of the admittance as a function of the
energies of the quantum dots' single-electron states. We model the the admittance
measured at the source electrode using a master equation approach similar to
Ref.~\cite{Cottet}. We take into account all relevant many body
states, i.e. for Fig.~3(c) and (d) the empty state and all one and two electron
states. For a given set of gate voltages $V_{L}$ , $V_{R}$  we
obtain the steady state density matrix at zero bias. We then
calculate the current flowing into the source electrode as a linear
response to a periodic driving of the source potential. This enables
us to deduce the double quantum dot admittance at the source electrode. We assume that the driving
frequency is too small to induce transitions between quantum dot
eigenstates and we therefore treat the perturbation as adiabatic.
The current into the source has two contributions: the particle
current due to tunneling from the source to the left dot and the
displacement current induced on the source capacitance by the
tunneling induced redistribution of charges on the quantum dot. We
include spin relaxation as well as phonon assisted tunneling between
the left and the right dot. Overall, we obtain good agreement with the
experimental results as demonstrated by Fig.~3.

\begin{figure}
\includegraphics[width=85mm]{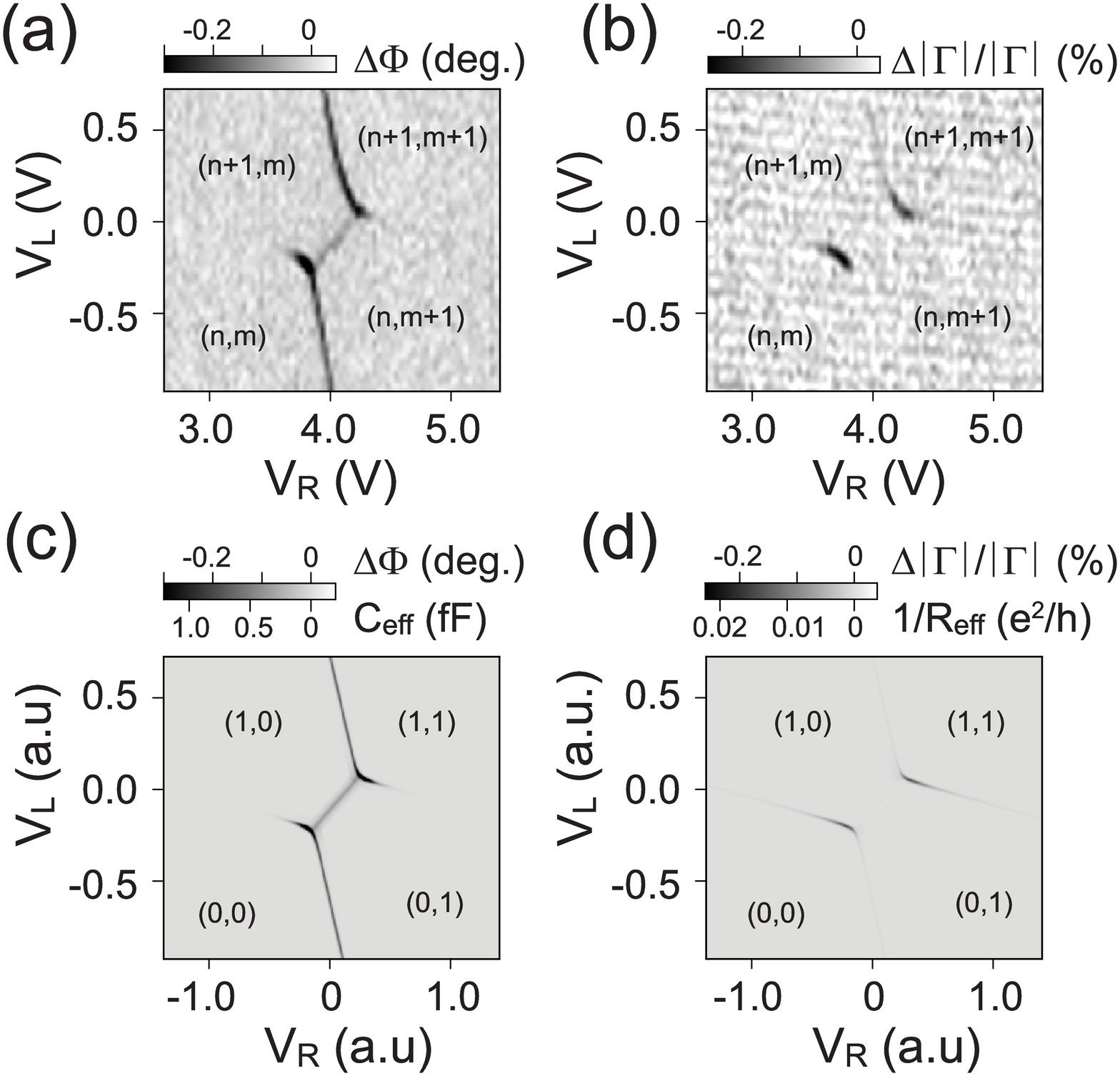}
\caption{\label{Fig3}\textbf{(a)} Measured phase shift of the carbon nanotube
double quantum dot device.\textbf{(b)} Measured amplitude response. \textbf{(c)} Calculated effective capacitance and phase shift of the double
quantum dot. The parameters used for the model calculations are $T=80$ mK. $U'=0.6$ meV, $t=40$ $\mu$eV, $\omega_0 = 1.2$ GHz, $\gamma_L = 0.15 \gamma_R$.
\textbf{(d)} Calculated conductance and damping for the same parameters as used in (c).}
\end{figure}

For a physical understanding of the results (and the model) it is instructive to consider several limiting cases.
Let us first consider transitions between $(n,m)$ and $(n,m+1)$
charge states such that direct
charge transfer between the two electrodes (a dc current) and
internal transitions of the double quantum dot can be neglected. An oscillating potential $V_{RF}(t)$ on
the right electrode modulates the energy difference $\delta\epsilon (t)$
between the right quantum dot states and the Fermi energy of the lead as $\delta\epsilon
(t) = \epsilon_0 -e \alpha V_{RF}(t)$, where $\epsilon_0$ is an offset,
tunable with the plunger gates. The constant $e\alpha$ converts
between voltage and energy and its value depends on the various
geometric capacitances of the device \cite{Wiel}. As a results of
the oscillating potential, charge moves back and forth between the
right electrode and right quantum dot. Depending on the ratio of the tunnel rate
$\gamma_R$ and angular driving frequency $\omega_0$, the induced
current has both in-phase and out-of-phase components with
respect to the voltage which can be expressed as a complex admittance
$Y(\omega_0)= R_{eff}^{-1} + j \omega_0 C_{eff}$. In the incoherent limit, i.e. for $h
\gamma_R \ll k_B T$ the terms are given by \cite{Gabelli}:
\begin{equation}\label{Reff}
    R_{eff}=\frac{4 k_B T}{e^2 \alpha^2 \gamma_R} \left(1 + \frac{\gamma_R^2}{\omega^2_0}\right)
\end{equation}
\begin{equation}\label{Ceff}
    C_{eff}=\frac{e^2 \alpha^2}{4 k_B T}\frac{1}{1+\frac{\omega^2_0}{\gamma_R^2}}
\end{equation}
As expressed by Eqs. (1) and (2), both the conductance and capacitance are dependent on the ratio of the tunnel coupling and angular driving frequency.
In the transparent limit ($\gamma \gg \omega_0$), the effective capacitance can be approximated by $C_{eff}
\approx e^2 \alpha^2 / 4 k_B T$. The conductance has a vanishing contribution, i.e. $R^{-1}_{eff}\rightarrow 0$, in both the
transparent and opaque limit. This can be understood intuitively as for very large $\gamma$ electrons will tunnel on the quantum dot as soon as
this is energetically possible and no energy is dissipated in the process. In the opaque limit, tunneling occurs out of equilibrium. However, as the probability of a tunnel event on the timescale of the driving frequency vanishes for weak coupling, no energy is dissipated either. Damping is therefore strongest in the intermediate regime where $\gamma \sim \omega$, as recently observed for single-electron tunneling devices coupled to electrical \cite{Persson, Ferguson} and mechanical \cite{Jun} resonators.

We can compare these predictions with the experimental data obtained on the carbon nanotube double quantum dot. Using $\alpha \approx 0.35$, as determined
from dc transport experiments, we obtain quantitative agreement between the experiment ($\Delta \Phi \sim 0.18$ degrees) and the change in capacitance predicted by Eq.~2 for the  $(n,m)$-$(n,m+1)$ transition, if we assume an effective
electron temperature $T \sim 80$ mK \cite{limit}. This is somewhat larger than the base temperature of the dilution fridge, most likely due to heat loading of the device by the coaxial cables. The estimate of the effective capacitance at the $(n,m)$-$(n+1,m)$ transition, i.e. due to charge transfer between the left lead and left dot also follows Eq.~2. However in this case, the prefactor is much smaller $\alpha \sim 0.05$ reflecting the weak coupling between the left quantum dot and the right electrode at which the rf signal is applied. Since the expected phase shift $\propto \alpha^2$ the response along this line is too weak to be detected. We observe a very weak amplitude modulation along the $(n,m)$-$(n,m+1)$ charge transition, $\Delta |\Gamma| /|\Gamma| \sim 0.05 \%$, most clearly observed in the top half of Fig.~3(b). This is consistent with damping due to out-of-equilibrium tunneling and in agreement with the theoretical calculations of the effective conductance shown in Fig.~3(d). The fact that the signal is rather weak implies that $\gamma_R \gtrsim \omega_0$ in our device. Stronger damping is observed at the triple points where the behavior of the device is similar to that of the conventional rf single-electron transistor \cite{Schoelkopf}.

Of particular interest is the phase signal along
the polarization line which reflects the movement of electrons
between the quantum dots, i.e. transitions between the $(n+1,m)$ and
$(n,m+1)$ charge states. In this case, the amplitude and width of
the signal is not set by temperature but the tunnel coupling $t$.
More precisely, it has been predicted \cite{Cottet} that, $C_{eff} = e^2 \beta^2/
4t$, where here $\beta \sim 0.6$ converts between $V_{RF}$ and detuning
$\epsilon_L - \epsilon_R$. The predictions can be directly verified with our experiments. The tunnel coupling can be deduced from the
stability diagram \cite{Graber}, yielding $t \sim 40$ $\mu$eV, and the estimate for $C_{eff} \approx 0.46$ fF therefore has no free parameters. This result is in
excellent agreement with the experimental data of Fig.~3(a) where we measure a phase shift of $\sim 0.11$ degrees, roughly a factor $\sim 2$
smaller than that observed at the $(n,m)-(n+1,m)$ transition, as also seen in the model calculation of Fig.~3(c).

We note that we did not observe spin blockade in dc transport experiments on the carbon nanotube double quantum dot studied here and the effect
is therefore not taken into account in the analysis. Nevertheless, spin blockade has been observed previously in carbon nanotubes \cite{Churchill1,Churchill2,Chorley}. In the
spin blockade regime, transitions between a (1,1) and (0,2) charge state directly depend on whether the (1,1) state is a singlet or triplet. Since the (0,2) charge state is a singlet
by virtue of the Pauli principle the (1,1) triplet state will be a blocked state which is reflected in the admittance \cite{Cottet}. Measurements of the admittance of a carbon nanotube double quantum dot as demonstrated here can therefore also be used for spin state readout \cite{Petersson}.

In conclusion, we have measured the complex admittance of a carbon nanotube double quantum
using rf reflectometry. The results are in quantitative agreement with a theoretical model of the device of which several limiting cases are discussed in detail. The demonstrated technique is of particular interest as a tool for fast and sensitive charge and spin state readout of carbon nanotube quantum dots. A further interesting possibility is to
extend the measurement to the coherent limit where the nanotubes are more strongly coupled to the leads. This should allow the
observation of a quantized charge relaxation resistance \cite{Gabelli} and possible deviations thereof in long nanotubes for which interactions
(Luttinger liquid physics) become important \cite{Martin,Burke}.

We thank David Cobden and Jiang Wei for the carbon nanotube growth and Andrew Ferguson for useful discussion.
This work was supported by the Newton Trust and the Royal Society
(M.R.B.).


\begin{thebibliography}{10}


\bibitem{Buttiker1}
M. B\"{u}ttiker, A. Pr\^{e}tre, and H. Thomas, Phys. Rev. Lett. \textbf{70},
4114 (1993).

\bibitem{Buttiker2}
A. Pr\^{e}tre, H. Thomas, and M. B\"{u}ttiker, Phys. Rev. B \textbf{54},
8130 (1996).

\bibitem{Loss}
D. Loss and D.P. DiVincenzo, Phys. Rev. A \textbf{57}, 120
(1998).

\bibitem{Cottet}
A. Cottet, C. Mora, and T. Kontos, Phys. Rev. B \textbf{83}, 121311
(2011).

\bibitem{Gabelli}
J. Gabelli, G. F\`{e}ve, J.-M. Berroer, B. Pla\c{c}ais, A. Cavanna,
B. Etienne, Y. Jin, and D.C. Glattli, Science \textbf{313}, 499
(2006).

\bibitem{Delbecq}
M.R. Delbecq, V. Schmitt, F.D. Parmentier, N. Roch, J.J. Viennot, G. F\`{e}ve, B. Huard,
C. Mora, A. Cottet, and T. Kontos, arXiv:1108.4371.

\bibitem{Petersson}
K.D. Petersson, C.G. Smith, D. Anderson, P. Atkinson, G.A.C. Jones,
and D.A. Ritchie, Nano Lett. \textbf{10}, 2789 (2010).

\bibitem{Kuemmeth}
F. Kuemmeth, S. Ilani, D.C. Ralph, and P.L. McEuen, Nature (London)
\textbf{452}, 448 (2008).

\bibitem{Galland}
C. Galland and A. Imamo\v{g}lu, Phys. Rev. Lett. \textbf{101},
157404 (2008).

\bibitem{Churchill1}
H.O.H. Churchill, A.J. Bestwick, J.W. Harlow, F. Kuemmeth, D. Marcos,
C.H. Stwertka, S.K. Watson, and C.M. Marcus, Nature Phys. \textbf{5}, 321 (2009).

\bibitem{Churchill2}
H.O.H. Churchill, F. Kuemmeth, J.W. Harlow, A.J. Bestwick, E.I. Rashba, K. Flensberg,
C.H. Stwertka, T. Taychatanapat, S.K. Watson, and C.M. Marcus, Phys. Rev. Lett. \textbf{102},
166802 (2009).

\bibitem{Jespersen}
T.S. Jespersen, K. Grove-Rasmussen, J. Paaske, K. Muraki, T.
Fujisawa, J. Nyg{\aa}rd, and K. Flensberg, Nature Phys. \textbf{7},
348 (2011).

\bibitem{Chorley}
S.J. Chorley, G. Giavaras, J. Wabnig, G.A.C. Jones, C.G. Smith,
G.A.D. Briggs, and M.R. Buitelaar, Phys. Rev. Lett. \textbf{106},
206801 (2011).

\bibitem{Biercuk}
M.J. Biercuk, D.J. Reilly, T.M. Buehler, V.C. Chan, J.M. Chow, R.G.
Clark, and C.M. Marcus, Phys. Rev. B \textbf{73}, 201402(R) (2006).

\bibitem{capacitance}
The main component of the capacitance is the parallel plate
capacitor formed by the bond pad and Si back gate.

\bibitem{Wiel}
W.G. van der Wiel, S. De Franceschi, J.M. Elzerman, T. Fujisawa, S.
Tarucha, and L.P. Kouwenhoven, Rev. Mod. Phys. \textbf{75}, 1
(2002).

\bibitem{Liang}
W.J. Liang, M. Bockrath, and H. Park, Phys. Rev. Lett. \textbf{88}, 126801 (2002).

\bibitem{Buitelaar}
M.R. Buitelaar, A. Bachtold, T. Nussbaumer, M. Iqbal and C. Sch\"{o}nenberger, Phys. Rev. Lett. \textbf{88},
156801 (2002).

\bibitem{Pozar}
D.M. Pozar, \textit{Microwave Engineering}, 3rd ed. (Wiley, New York, 2005).

\bibitem{Persson}
F. Persson, C.M. Wilson, M. Sandberg, G. Johansson, and P.
Delsing, Nano Lett. \textbf{10}, 953 (2010).

\bibitem{Ferguson}
C. Ciccarelli and A.J. Ferguson, arXiv:1108.3463.

\bibitem{Jun}
Z. Jun, M. Brink, P.L. McEuen, Nano Lett. \textbf{8}, 2399 (2008).

\bibitem{limit}
For this temperature $k_B T / h \gamma_R \sim 2$.

\bibitem{Schoelkopf}
R.J. Schoelkopf, P. Wahlgren, A.A. Kozhevnikov, P. Delsing, and D.E.
Prober, Science \textbf{280}, 1238 (1998).

\bibitem{Graber}
M.R. Gr\"{a}ber, W.A. Coish, C. Hoffmann, M. Weiss, J. Furer, S.
Oberholzer, D. Loss, and C. Sch\"{o}nenberger, Phys. Rev. B
\textbf{74}, 075427 (2006).

\bibitem{Martin}
Y. Hamamoto, T. Jonckheere, T. Kato, and T. Martin, Phys. Rev. B
\textbf{81}, 153305 (2010).

\bibitem{Burke}
P.J. Burke, IEEE T. Nanotechnol. \textbf{1}, 129 (2002).

\end{thebibliography}
\end{document}